\documentclass[pra,showpacs,preprintnumbers,tightenlines,superscriptaddress,twocolumn]{revtex4}
\usepackage{dcolumn}
\usepackage{bm}
\usepackage{epsfig}
\usepackage{stmaryrd}
\usepackage{dsfont}
\usepackage{bbm}
\usepackage{amssymb}
\usepackage{amscd}
\usepackage{amstext}
\usepackage{rotating,standalone}
\usepackage{graphicx}
\usepackage{physics}
\usepackage{color}
\usepackage[version=4]{mhchem}
\usepackage{relsize}
\usepackage{amsmath}
\usepackage{float}
\usepackage{amsfonts}
\usepackage{hyperref}

\newcommand{\bras}[1]{\langle\,{#1}\, |}
\newcommand{\kets}[1]{|\,{#1}\,\rangle}
\newcommand{\rvec}{\mathbf{r}}

\newcommand{\id}{\mathds{1}}
\newcommand{\sub}[2]{{#1}_{\mbox{\!\! \scriptsize #2}}}

\def\beq{\begin{equation}}
\def\eeq{\end{equation}}

\def\CR{\nonumber\\[0.15cm]}

\newcommand{\fref}[1]{Fig.~\ref{#1}}

\newcommand{\eref}[1]{Eq.~(\ref{#1})}

\newcommand{\sref}[1]{section~\ref{#1}}

\newcommand{\cref}[1]{chapter~\ref{#1}}

\newcommand{\Cref}[1]{Chapter~\ref{#1}}

\newcommand{\bref}[1]{(\ref{#1})}

\begin{document}
\title{Multi-Excitons in Flexible Rydberg Aggregates}
\author{G.~Abumwis}
\affiliation{Department of Physics, Bilkent University, Ankara 06800, Turkey}
\affiliation{Max Planck Institute for the Physics of Complex Systems, N\"othnitzer Strasse 38, 01187 Dresden, Germany}
\author{S.~W\"uster}
\affiliation{Department of Physics, Bilkent University, Ankara 06800, Turkey}
\affiliation{Department of Physics, Indian Institute of Science Education and Research (IISER) Bhopal, Madhya Pradesh, 462066, India}
\email{sebastian@iiserb.ac.in}
\begin{abstract}
Flexible Rydberg aggregates, assemblies of few Rydberg atoms coherently sharing electronic excitations while undergoing directed atomic motion, show great promise as quantum simulation platform for nuclear motion in molecules or quantum energy transport. Here we study additional features that are enabled by the presence of more than a single electronic excitation, thus considering multi-exciton states. We describe cases where these can be decomposed into underlying single exciton states and then present dynamical scenarios with atomic motion that illustrate exciton-exciton collisions, exciton routing, and strong non-adiabatic effects in simple one-dimensional settings.
\end{abstract}
\maketitle

\section{Introduction}

Ultra-cold Rydberg atoms are now recognised as a promising tool for scalable neutral atom quantum information \cite{Saffman:quantinfryd:review}, quantum non-linear optics \cite{Mohapatra:giantelectroopt,sevincli:nonlocopt,peyronel:quantnonlinopt,dudin:singlephotsource} or the simulation of condensed matter systems \cite{Weimer-Buchler-Rydbergquantumsimulator-2010,Zeiher:ryd_dress_spinchain,Glaetzle:spinice:prx,Marcuzzi:manybodyloc}.
In such applications, the thermal or interaction-induced motion of atoms gives rise to detrimental but unavoidable noise sources \cite{wilk:entangletwo,mueller:browaeys:gateoptimise}. This problem is turned into an asset when using Rydberg atoms for the quantum simulation of nuclear dynamics in complex molecules, within flexible Rydberg aggregates \cite{wuester:review}, where Rydberg atomic motion replaces nuclear motion albeit on much larger and more accessible length scales.

If desired, Rydberg atoms can be easily put into motion, either by van-der-Waals interactions or the more interesting dipole-dipole interactions \cite{book:gallagher,noordam:interactions}. This motion is then governed by Born-Oppenheimer (BO) surfaces that lead to dynamics of repulsive, attractive or indeterminate nature, depending on the overall electronic state of the Rydberg system \cite{cenap:motion,wuester:review}. Earlier work has shown interesting interplay between excitation transport and this atomic motion even in simple one-dimensional settings \cite{wuester:cradle,moebius:cradle,zoubi:VdWagg}.

These studies have so far been restricted to cases where a single excitation in an angular momentum $p$ ($l=1$)-state is undergoing transport on a backbone of Rydberg atoms that are otherwise in $s$ ($l=0$).
Here we present initial investigations of the more complex case, where multiple $p$-excitations are present. This will lead to a larger variety of dipole-dipole BO surfaces that can be designed. While the experimental excitation of Rydberg aggregates with precisely one excitation is in principle straightforward \cite{moebius:cradle,wuester:cannon}, there will always be a small probability of multiple excitation that further motivates our studies.
Additionally, multi-exciton states have been thoroughly studied in molecular aggregates, especially in light harvesting complexes \cite{VANGRONDELLE19941,PhysRevLett.78.3406}, thus by accessing multi-exciton states, we widen the scope of Rydberg aggregates for the quantum simulation of light harvesting \cite{schoenleber:immag}. 

We show that multi-exciton scenarios can be treated formally in a similar way to those containing a single exciton, initially demonstrating how for specific cases the spectrum of two-exciton states can be obtained by a decomposition into single-exciton states. However in general, this decomposition will be limited. When considering atomic motion due to multi-excitonic BO surfaces, we focus on several examples that highlight key scenarios beyond those explorable within the single excitation manifold, such as: interaction of two excitation pulses, excitation transport switching and strong non-adiabatic effects for one-dimensional motion.

This article is organized as follows:  
In \sref{sec2}, we introduce a theoretical description of bi-excitons in one-dimensional Rydberg aggregates. Next, in \sref{excitonstates} we compare bi-excitons with excitons for two cases and show when bi-excitons can be written as product states of excitons. Then, in \sref{sec4}, we proceed to discuss the induced motion of atoms in the aggregate and its dependence on the initial state. In \sref{flexagg} we look into the mutual interplay between the motional and electronic dynamics, where we highlight a number of emerging scearios that are exclusive to bi-excitons. We show that two exciton pulses undergo elastic repulsive collisions (\sref{sec41}), illustrate the possibility of switching or gating one exciton pulse by another (\sref{sec42}) and finally identify a special case with particularly prominent effects of non-adiabatic transitions on the dynamics (\sref{sec43}).

\section{Multi-excitons} \label{sec2}

%
\begin{figure}[htb]
\centering
\epsfig{file=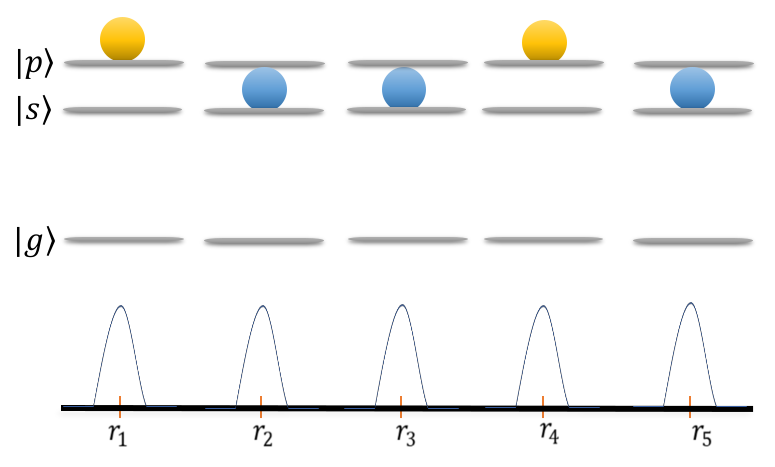,width= 0.99\columnwidth}
\caption{(top) Essential state diagram of a Rydberg aggregate, illustrating the state $\ket{\pi_{14}}$, see text. (bottom) Corresponding atomic positions and position uncertainties.
\label{sketch}}
\end{figure}
 We consider a chain of N identical Rydberg atoms that are confined to move in one dimension, with atomic positions described by $\textbf{R}=(\mathbf{r}_{1},\mathbf{r}_{2},\mathbf{r}_{3} ....)$ where $\mathbf{r}_{n}$ denotes the position of the \textit{n}th atom, see \fref{sketch}.  Each atom on the chain is initially prepared in one of two energetically close Rydberg states, a lower state $\ket{s}$ with energy $\varepsilon_{s}$ or an upper state $\ket{p}$ with $\varepsilon_{p}$ such that $\varepsilon_{p} > \varepsilon_{s}$. 
 The single exciton manifold with exactly one atom in a $p$ state and all others in the $s$ state is then spanned by the basis $\ket{\pi_n}=\ket{n}=\ket{sss\cdots p \cdots sss}$, where in state $\ket{\pi_n}$ only the $n$th atom is in $p$. To generalize the notation to $q$ excitations, we employ the basis $\kets{\Pi^{(q)}_j}=\ket{n_j m_j \cdots l_j}$, where $\{n_j,m_j,\cdots l_j\}$ are a set of $\bar{N}=\binom{N}{q}$ integers that indicate which $q$ atoms carry the $p$-excitations. The index $j$ is just numbering the possible combinations, and thus runs from $1\cdots  \bar{N}$. For example, in the case of two $p$ excitations we have $|\Pi_{1}^{(2)}\rangle=\ket{1,2}, $ $|\Pi_{2}^{(2)}\rangle=\ket{1,3}$ to $|\Pi_{10}^{(2)}\rangle=\ket{4,5}$. 
 
\subsection{Hamiltonian}

The Hamiltonian constrained to the single exciton state space is
   \begin{equation} 
   \mathcal{H}_{el}^{(1)}(\textbf{R})=\sum^{N}_{n,m:n\neq m} \dfrac{C_3}{R^{3}_{nm}}\ket{\pi_{n}}\bra{\pi_{m}}- \id\sum^{N}_{l,k:l\neq k}\dfrac{C_{6}}{2R^{6}_{lk}}.
   \label{H1p}
   \end{equation}
with $R_{nm}=|\rvec_{n}-\rvec_{m}|$ the distance between atoms $n$ and $m$. The operator $\id=\sum^{N}_{n}\ket{\pi_{n}}\bra{\pi_{n}}$ is the identity in electronic space.
 $C_3$ and $C_6$ are the dispersion coefficients for resonant dipole-dipole and Van-der-Waals (VdW) interactions respectively, where for the latter we assume equal interactions for $ss$ and $sp$ pair-states for simplicity.

In this article we will mainly focus on the case of exactly two $p$-excitations for which we can write:
  \begin{equation}
  \mathcal{H}_{el}^{(2)}(\textbf{R})=\sum^{\bar{N}}_{i,j:i\neq j}V_{ij}(\textbf{R}) \kets{\Pi^{(2)}_{i}}\bras{\Pi^{(2)}_{j}}-\id \mathlarger{\sum^{N}_{l,k:l\neq k}}\dfrac{C_{6}}{2R^{6}_{lk}},
   \label{H2p}
  \end{equation}
 where $\bar{N}=\binom{N}{2}$, $ \kets{\Pi^{(2)}_{i}}=\ket{n_i m_i}$ and,
 \begin{align}
 V_{ij}(\textbf{R})=&\bigg(\delta_{m_{i}m_{j}}\dfrac{C_3}{R^{3}_{n_{i}n_{j}}}+\delta_{n_{i}n_{j}}\dfrac{C_3}{R^{3}_{m_{i}m_{j}}} \CR
 &+\delta_{m_{i}n_{j}}\dfrac{C_3}{R^{3}_{n_{i}m_{j}}}+\delta_{n_{i}m_{j}}\dfrac{C_3}{R^{3}_{m_{i}n_{j}}}\bigg).
 \end{align}
The notation could be straightforwardly generalized to more than two excitations ($q>2$), but this will not be needed here.

\subsection{Excitons}

We call the eigenstates of the interacting Hamiltonians above (Frenkel) excitons \cite{frenkel:footnote}.
For later use, we will distinguish solutions of the single-excitation eigenproblem
\begin{equation} 
\label{eq_eig1}
  \mathcal{H}_{el}^{(1)}\ket{\varphi^{N}_{k}(\textbf{R})}=U_{k}(\textbf{R})\ket{\varphi^{N}_{k}(\textbf{R})},
 \end{equation}
from those for two excitations
  \begin{equation}
\label{eq_eig2}
    \mathcal{H}_{el}^{(2)}\ket{\zeta^{N}_{k}(\textbf{R})}=O_{k}(\textbf{R})\ket{\zeta^{N}_{k}(\textbf{R})}.
 \end{equation}
We keep the number of Rydberg atoms $N$ as an additional labelling parameter and $U_{k}(\textbf{R})$ ($O_{k}(\textbf{R})$) are the exciton energies 
(Born-Oppenheimer surfaces \cite{wuester:review}) for the one (two) excitation case. Accordingly $\ket{\varphi^{N}_{k}(\textbf{R})}$ ($\ket{\zeta^{N}_{k}(\textbf{R})}$) are the exciton eigen-states.

\section{Bi-exciton versus single exciton states} 
\label{excitonstates}

We will first contrast bi-excitons states with single exciton states for two simple cases: (i) A regular chain with equal distance between neighboring atoms set at $d=5 \mu m$, as shown in Fig. \ref{sketch}, and (ii) 
A chain with a dislocation at one end, where the final atom is distance of only $a=d/2$ from its neighbor. For both cases we solve \eref{eq_eig2} and refer to \cite{cenap:motion} for the solutions of  \eref{eq_eig1}.

\begin{figure}[htb]
	\centering
	\epsfig{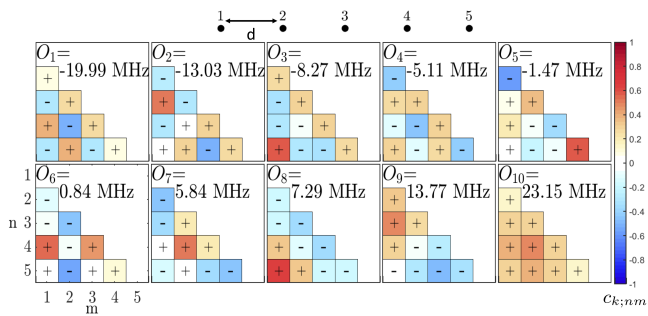} 
	\caption{Representation of all bi-exciton states $\zeta_k$ for case (i), a homogenous chain with separation $d=5 \mu m$ of five Rydberg atoms ($N=5$) with $q=2$ excitations. The atomic locations are sketched at the top. The color indicates the amplitude $c_{k;nm}$ (see text), and the row (column) of the tiles the indices $n$ ($m$) of $p$-excited atoms. 
		\label{homog_excitons}}
\end{figure}
These bi-exciton states have the structure $\ket{\zeta^{5}_{k}(\textbf{R})}=\sum c_{k;nm} \ket{nm}$, and we show the coefficients $c_{k;nm} = \braket{nm}{\zeta^{5}_{k}(\textbf{R})}$ for their visualisation.
For the homogenous chain [case (i)] these are shown in \fref{homog_excitons}. As required by symmetry all modes are either symmetric or anti-symmetric about the centre of the chain, and share the properties that the excitations are fully de-localized over the chain. For the illustrative exciton energies shown we assumed $C_3=976$ MHz $\mu m^3$, $C_6=5400$ MHz $\mu m^6$, roughly corresponding to Rydberg states with a principal quantum number $\nu\approx40$. At this separation, VdW interactions play a minor role for exciton states.

\begin{figure}[htbp]
\centering
\epsfig{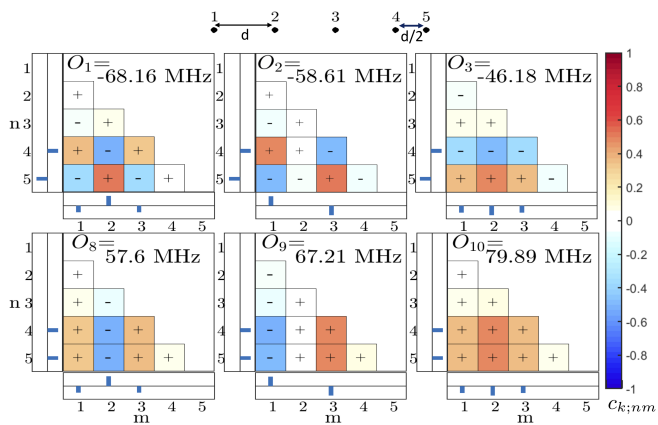} 
\caption{Decomposition of bi-excitons into products of single exciton states for case (ii), the dislocated chain with of five Rydberg atoms ($N=5$). The atomic locations are sketched at the top, the base separation is $d=5 \mu m$ and the dislocation  $a=2.5 \mu m$. 
We show the six cases that are a simple product of two single exciton states. The color indicates the amplitude $c_{k;nm}$ (see text), and the row (column) of the tiles the indices $n$ ($m$) of $p$-excited atoms.
Bordering the tiles we show the exciton state on the dislocation (atoms $4$, $5$, $y$-axis) and the remainder of the chain (atoms $1$-$3$, $x$-axis), where length and direction of the stick qualitatively indicate $c_{k;n}$, see text.
\label{decomposition}}
\end{figure}
It is easier to interpret bi-exciton states and link them to single exciton states for the dislocated chain [case (ii)] shown in \fref{decomposition}. In this case, the dipole-dipole interaction strength
between the two atoms in the dislocation, given by $C_3/|\mathbf{r}_4 - \mathbf{r}_5|^3$, provides by far the largest matrix elements in the Hamiltonian. Because of this, six of the states factor into simple tensor products of one single-exciton state for the dislocation and a second for the remainder of the chain. Specifically we find $\kets{\zeta^{(5)}_1} = \kets{\varphi^{(3)}_1}\otimes\kets{\varphi^{(2)}_1}$, $\kets{\zeta^{(5)}_2} = \kets{\varphi^{(3)}_2}\otimes\kets{\varphi^{(2)}_1}$, $\kets{\zeta^{(5)}_3} = \kets{\varphi^{(3)}_3}\otimes\kets{\varphi^{(2)}_1}$, $\kets{\zeta^{(5)}_8} = \kets{\varphi^{(3)}_1}\otimes\kets{\varphi^{(2)}_2}$, $\kets{\zeta^{(5)}_9} = \kets{\varphi^{(3)}_2}\otimes\kets{\varphi^{(2)}_2}$, $\kets{\zeta^{(5)}_{10}} = \kets{\varphi^{(3)}_3}\otimes\kets{\varphi^{(2)}_2}$, where the single exciton states have the structure $\kets{\varphi_{k}}=\sum c_{k;n} \ket{n}$ and the coefficients $c_{k;n}$ are visualized as bars in \fref{decomposition}.  In the list above the state to the left of the tensor product $\otimes$ pertains to the atoms $1$-$3$ (chain) and the one to the right to the atoms $4$-$5$ (dislocation). To give one example of this tensor product notation, the state $\kets{\zeta^{(5)}_1}$ is very well approximated by
\begin{align}
\kets{\zeta^{(5)}_1} &= - \kets{\varphi^{(3)}_1}\otimes\kets{\varphi^{(2)}_1}\\
&= \big(\frac{1}{2}\ket{\pi_1} - \frac{1}{\sqrt{2}}\ket{\pi_2} + \frac{1}{2}\ket{\pi_3}  \big)\otimes \frac{1}{\sqrt{2}} \big(\ket{\pi_4} - \ket{\pi_5} \big)\CR
&=  \frac{1}{2\sqrt{2}} \big( \ket{14} -\sqrt{2} \ket{24} + \ket{34} - \ket{15} + \sqrt{2} \ket{25} - \ket{35}\big).\nonumber
\end{align}
In the last line we used the states $\ket{\Pi^{(2)}}$ defined in \sref{sec2}.

\begin{figure}[htb]
\centering
\epsfig{file=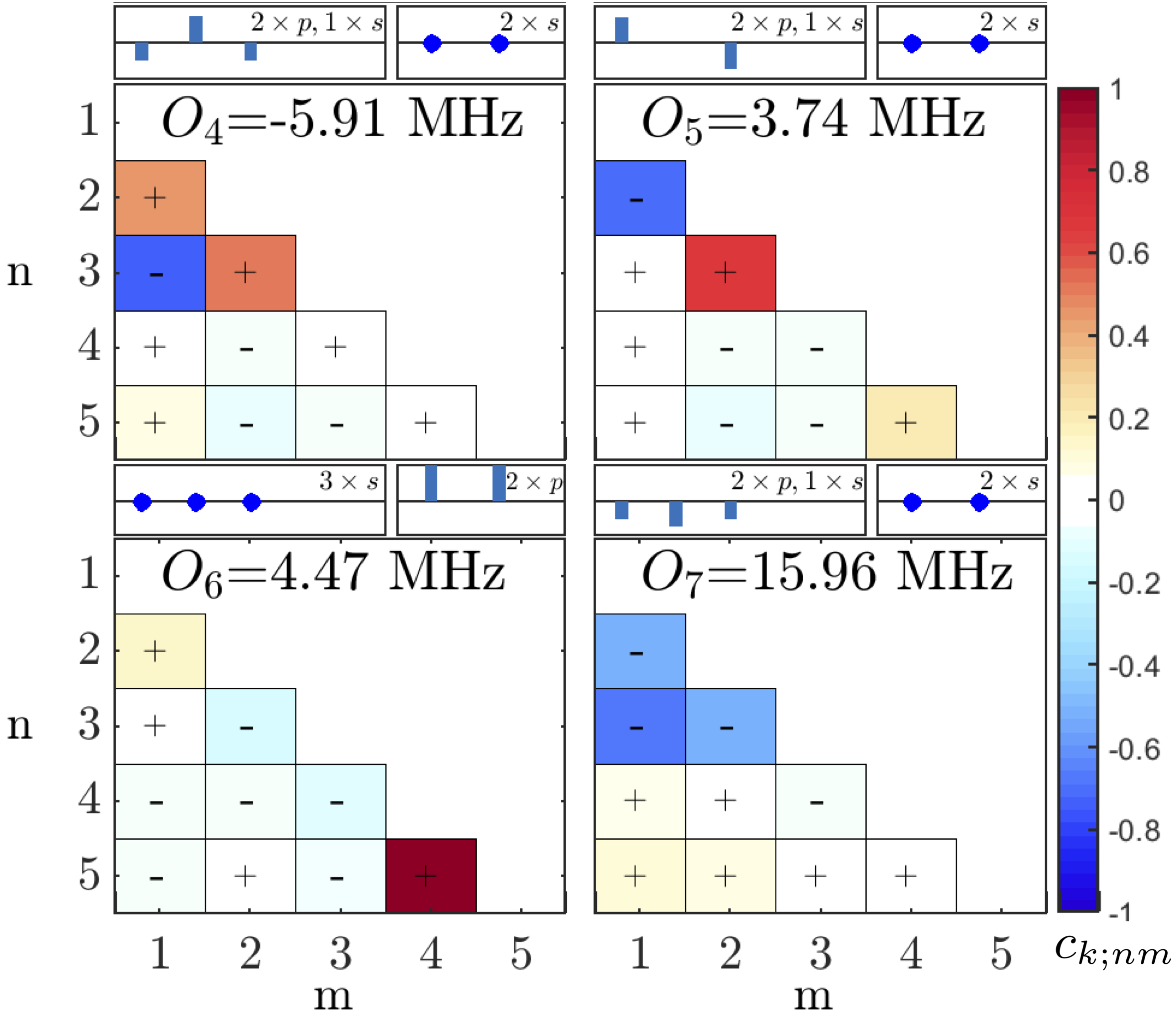,width= 0.9\columnwidth} 
\caption{Special bi-exciton states that cannot be de-composed into single exciton ones. The labelling is as in \fref{homog_excitons} and \fref{decomposition}.
\label{newstates}}
\end{figure}

There are also four states that cannot be reduced to known single exciton results, shown in \fref{newstates}. These can be understood as the state with both $p$-excitations on the dislocation, and three states where both $p$-excitations avoid the dislocation and reside on the main chain. In the latter three cases, these then form essentially single exciton state on the main chain, for which the role of the $s$ and $p$ electronic levels are inverted. These can be written in a basis $\kets{\bar{\pi}_n}=\ket{ppp\cdots s \cdots pppp}$, where in this state all atoms are in $p$ except the $n$'th one which is in $s$. Writing the corresponding inverted exciton states as $\kets{\bar{\varphi}_k}$, we can write $\kets{\zeta^{(5)}_4} = \kets{\bar{\varphi}^{(3)}_1}\otimes\kets{ss}$, $\kets{\zeta^{(5)}_5} = \kets{\bar{\varphi}^{(3)}_2}\otimes\kets{ss}$, $\kets{\zeta^{(5)}_7} = \kets{\bar{\varphi}^{(3)}_3}\otimes\kets{ss}$ and $\kets{\zeta^{(5)}_6} = \kets{sss}\otimes\kets{pp}$.
%

\section{Atomic motion in a flexible Rydberg aggregate} \label{sec4}

The dipole-dipole induced motion of atoms in a flexible bi-exciton aggregate could be studied by solving the time-dependent Schr{\"o}dinger equation
\begin{equation} \label{eq_TDS}
i\hbar \frac{\partial}{\partial t} \Psi(\textbf{R},t)= \left(-\frac{\hbar^2}{2m}\sum_{n=1}^N \boldsymbol{\nabla}^2_n +\mathcal{H}_{el}^{(q)}(\textbf{R}) \right)\Psi(\textbf{R},t),
\end{equation}
for atoms of mass $m$. Even when restricting the motion of atoms to one-dimension, as we will do in the following, the direct solution of \eref{eq_TDS} for $N=5$ would be too challenging.
Hence we resort to quantum-classical propagation, where the motion of atoms is treated classically. This works very well for flexible Rydberg aggregates \cite{wuester:review,wuester:cradle,leonhardt:switch}.

Expanding the electronic quantum states as $\ket{\Psi(t)}=\sum c_{nm}(t) \ket{nm}$, we thus solve
\begin{equation}
\label{elec_explicit}
i\hbar \frac{\partial}{\partial t} c_{nm}(t) = \sum_{kl} \bra{nm} \mathcal{H}_{el}^{(q)}(\textbf{R}) \ket{kl} c_{kl}(t),
\end{equation}
which is coupled to the motion of atoms 
\begin{equation}
\label{newton}
\textbf{F}=M\ddot{\textbf{R}}=-\grad_{\textbf{R}}O_{\gamma(t)}(\textbf{R}).
\end{equation} 
Here $\gamma(t)$ is a stochastically varied index that describes on which BO surface the system is currently evolving. The stochastic jumps between surfaces correspond to non-adiabatic transitions between exciton states. The likelihood of a transitions from surface $i$ to surface $k$ is set by the non-adiabatic coupling vector $\textbf{d}_{ki}=\braket{\zeta^{(N)}_k}{\grad_R\phi^{(N)}_i}$. All results shown are then averaged over a set of 
$\sub{N}{traj}$ trajectories, typically starting from a chosen initial BO surface, i.e.~$\gamma(t=0)=\sub{k}{ini}$.

For further details on this Tully's fewest switching algorithm, we refer to the original chemical physics literature \cite{tully:hopping,tully:hopping2,tully:hopping:veloadjust}, the review \cite{barbatti:review_tully}
or our earlier work \cite{moebius:cradle,leonhardt:orthogonal}.

\begin{figure}[htb]
\centering
\epsfig{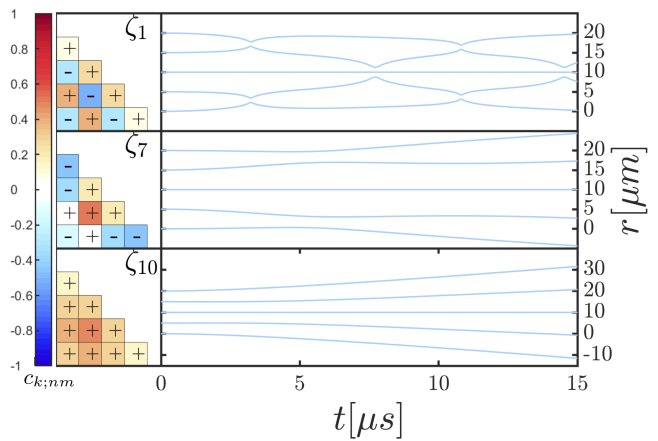} 
\caption{Repulsive versus attractive atomic motion for selected bi-exciton states. The electronic aggregate states are shown on the left, in the same style as for \fref{homog_excitons} and are forced to be constant in time. The right panels show the evolving positions of the five atoms under the influence of the force corresponding to the Born-Oppenheimer surface that corresponds to the aggregate state in the same row on the left.
\label{trajectories}}
\end{figure}

To obtain a first idea of the allocation of motional dynamics to exciton state, we can proceed as in \cite{cenap:motion}, and initially ignore the evolution of the electronic state \bref{elec_explicit}, fix $\gamma=k$ as constant in time, and thus just obtain the Newtonian motion on a fixed BO surface as shown in \fref{trajectories}. We see that, as for single-excitation aggregates, the type of motion is set by the overall aggregate electronic states.
As usual opposite amplitude signs between neighboring atoms result in attractive dynamics, while the same sign results in repulsive dynamics (for $C_3>0$). The bounces for the attractive case result from repulsive van-der-Waals interactions at very short distances, that are independent of the exciton state. VdW interactions thus preclude a likely ionising close encounter of Rydberg atoms, which is why we included them in the model. The parameters for these simulations and the following ones are as listed in \sref{excitonstates}, for $^7$Li atoms of mass {$10^{-26}$ Kg}. 

\section{Interplay of motional and electronic dynamics}
\label{flexagg}

For a more complete picture, we now proceed to treat the coupled electronic and motional dynamics of the flexible Rydberg aggregate, also taking into account atomic position uncertainties and non-adiabatic transitions.
To that end, we consider a system that is prepared in one of the electronic Hamiltonian eigenstates $\ket{\zeta^N_k}$, with atoms assumed to be initially optically trapped in a harmonic potential before they are released to freely move in one dimension. We thus use a Gaussian probability distribution for initial positions with a standard deviation $\sigma$ (the ground-state width of the trap) and the same for initial velocities with width $\sigma_v=\hbar/(M\sigma)$. Our results are then averaged over a large number $\sub{N}{traj}$ of trajectories.  For the following numerical simulations, we used  $\sigma=0.3$ $\mu$m and propagated $\sub{N}{traj}=10^4$ trajectories. As before, the main distance between nearest neighbours is $d=5 \mu m$; when a dislocation is present, the distance between the two atoms forming the dislocation was $a=d/2$.

In the following we present several selected dynamical scenarios, in which the presence of a second p-excitation causes the emergence of crucial additional features, compared to similar scenarios with just a single p-excitation.

\subsection{Interactions of Exciton pulses} \label{sec41}

In earlier work we had shown that in a flexible Rydberg aggregate, a pulse of atomic dislocation in a regular chain, with ensuing fast motion, combined with electronic excitation exhibits particularly interesting transport properties. We had called these "exciton(-motion) pulses" in \cite{leonhardt:switch,leonhardt:orthogonal}. It can adiabatically deliver a shared excitation and the associated entanglement from one end of a chain to the other \cite{wuester:cradle}. This bears some resemblance to a Davydov soliton \cite{davydov:soliton,weidlich:pulse}, albeit without the restoring force prominently part of the latter. These pulses can then be split and their coherence properties controlled, e.g.~by modifications of chain geometry or conical intersections \cite{leonhardt:switch,leonhardt:orthogonal}.

\begin{figure}[htb]
\centering
\epsfig{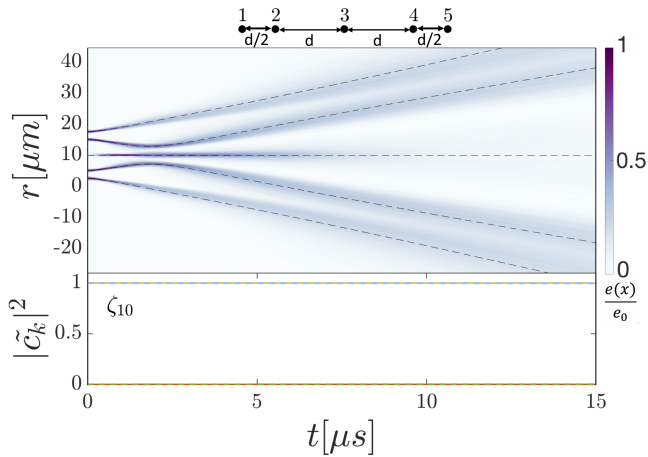} 
\caption{Elastic collision of two excitation carrying pulses of atomic motion. (top) Mean positions of the atoms (dashed black line) on top of the weighted excitation density $e(r,t)$ (color shade). The latter represents a spatial binning of all atomic positions, each weighted with the probability to carry any excitation, see also \cite{leonhardt:orthogonal}. With $e_0$ we denote $\mbox{max}_{r,t} e(r,t)$. (bottom) Adiabatic populations $|\tilde{c}_k|^2$, where $\tilde{c}_k(t) = \braket{\zeta_k(\mathbf{R}(t))}{\Psi(t)}$. Only $|\tilde{c}_{10}|^2$ for the initial repulsive surface is visibly non-zero, indicating essentially adiabatic dynamics. The initial repulsive state has the same character as $\ket{\zeta_{10}}$ in \fref{homog_excitons}, but with suppressed population on the centre atom.
\label{interactions}}
\end{figure}
Now, with two p-excitations, we are in a position to study the interaction of two such exciton motion pulses. These are shown in \fref{interactions}, for a scenario where each excitation is localized at a different end of a chain, with a dislocation on both ends. As the system evolves the excitation pulses collide in the chain centre, due to both pulses having repulsive character. This means that on each dislocation (top and bottom), one p-excitation is shared by the two atoms $\sim\ket{sp}+\ket{ps}$. The resultant contribution to the potential energy is then positive. Finally, when colliding at the centre, repulsive vdW interactions are at play and the pules reflect off each other. The figure shows the weighted excitation density, showing that all excitation is reflected off the centre. The second panel indicates that the initial electronic state is being adiabatically followed with high fidelity.

\subsection{Gate for exciton pulses} \label{sec42}

When decomposed into single excitation states, the scenario in the previous section amounts to the collision of two exciton states of the same type, connected to the repulsive dimer state on the dislocation \cite{cenap:motion}. We now show that one exciton pulse impacting onto a chain occupied by a second excitation may be routed through the chain, or reflected off it, dependent on the exciton state realized on the chain. 

\begin{figure}[htb]
\centering
\epsfig{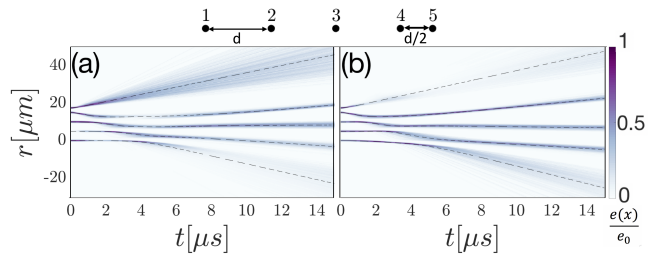} 
\caption{Transmission gate for exciton-motion pulses. An incoming exciton motion pulse (first exciton) is reflected or transmitted depending on the exciton state of the target chain (second exciton). (a) Reflection for gate exciton in $\kets{\varphi^{(5)}_2}$. (b) Transmission for gate exciton in $\kets{\varphi^{(5)}_3}$.
\label{switch}
}
\end{figure}

This is shown in \fref{switch}, where the dynamics in (a) is initialized in state $\kets{\zeta^{(5)}_9}$ while (b) commences from $\kets{\zeta^{(5)}_{10}}$. Both have in common that an incoming exciton-motion pulse is associated with the dislocated atoms $1$, $2$ sharing one $p$-excitation, while the second $p$ is distributed over atoms $3-5$. 

According to our decomposition in \sref{excitonstates} the non-dislocated part of the chain is thus in the (single $p$) exciton state $\kets{\varphi^{(5)}_2}$ in (a) and $\kets{\varphi^{(5)}_3}$ in (b). We can see that in (a) the incoming excitation is reflected off the chain, finally residing on the atom going out towards positive $r$, while the incoming momentum is transferred through as usual \cite{wuester:cradle,moebius:cradle}. In contrast for case (b), the excitation is passed through the chain together with the momentum pulse and finally resides on the
atom moving out towards negative $r$.

While our quantum-classical approach is not able to show this, the conditional passage should even be quantum coherent if the motional wave function of the Rydberg aggregate remains quantum coherent.
Note that the full-system life-time for our parameters would be $\sub{\tau}{agg}=(2/\tau_p + 3/\tau_s)^{-1}=19.4$ ${\mu}$s for $\tau_s=70$ ${\mu}$s and $\tau_p=232$ ${\mu}$s due to spontaneous decay and black-body redistribution at $T=300K$ \cite{beterov:BBR}. Hence we expect the motional dynamics shown to take place before the first decay of a Rydberg atom involved is to be expected.

\subsection{Non-adiabatic transitions} \label{sec43}

Finally, we show a scenario that leads to clear and significant non-adiabatic effects even within a \emph{one-dimensional} flexible Rydberg aggregate. Earlier, non-adiabatic effects were most prominently reported only once motion of each atom was taking place in two-dimensions \cite{wuester:CI,leonhardt:switch,leonhardt:orthogonal,leonhardt:unconstrained}.

\begin{figure}[htb]
\centering
\epsfig{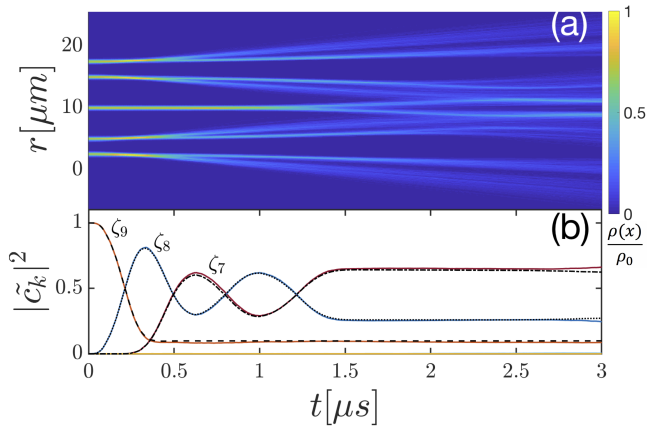} 
\caption{Non-adiabatic evolution of a one-dimensional flexible Rydberg aggregate with two excitons. (a) Total atomic density $\rho(x,t)$ (binned atom positions) as a function of time. With $\rho_0$ we denote $\mbox{max}_{r,t} \rho(r,t)$. Multiple spatial modes are clearly visible. (b) Adiabatic populations $\tilde{p}_k=|\tilde{c}_k|^2$ (colored lines) and matching trajectory fractions (dashed lines), indicating strong non-adiabatic effects.
\label{nonadiabdyn}}
\end{figure}

Non-adiabatic effects can be seen in \fref{nonadiabdyn} when starting from an electronic state akin to $\kets{\zeta^5_9}$, albeit for a doubly dislocated chain. We notice in the total atomic density (not excitation density) shown in panel (a) that the densities for each atom start splitting around $0.5\mu s$, indicating a superposition occupying simultaneously multiple BO-surfaces with consequently disparate forces acting on atoms.
This is corroborated by the adiabatic populations $\tilde{p}_k=|\tilde{c}_k|^2$, defined in the caption of \fref{interactions} and shown here in panel (b). In contrast to the earlier scenarios, these now show large variation with ultimately three BO surfaces strongly involved in the dynamics. The figure also includes the typical consistency check for the employed quantum classical surface hopping algorithm, comparing the adiabatic populations $\tilde{p}_k(t)$ with the fraction $f_k(t)$ of trajectories currently evolving on the matching BO surface. Here $f_k(t)$ is defined as 
\begin{equation}
f_k(t)=\frac{1}{N_{traj}}\sum_{i=1}^{N_{traj}}\delta_{k\gamma(t)}, 
\end{equation}  
where $k$ is a chosen adiabatic surface and $\gamma(t)$ represents the adiabatic surface that the system follows as it evolves in time, see \eref{newton}. 
We can see that surface hopping is consistent (indicating the absence of forbidden jumps), and that signatures of non-adiabaticity are visible in atomic density profiles even within this simple one-dimensional scenario, if density profiles can be measured with about $2\mu$m spatial resolution.

\section{Conclusions and outlook}

We have extended earlier work on flexible Rydberg aggregates from the single excitation manifold to the two exciton manifold. After presenting the decomposition of new exciton states into those known earlier, we highlighted several dynamical features that newly arise due to the addition of the second excitation. These were collisions of exciton-motion pulses, exciton pulse gates or routers controlled by the exciton state of a gate chain, as well as the occurrence of strong and clear non-adiabatic effects in scenarios with one-dimensional motion.

Most of the scenarios shown here represent examples of joint quantum transport and electronic excitation transport that might inspire novel transport processes in devices or artificial light harvesting \cite{saikin:excitonreview}. Multi-exciton processes are of relevance in light harvesting \cite{VANGRONDELLE19941,PhysRevLett.78.3406}, particularly when exciton-exciton annihilation processes are involved \cite{CAMPILLO197693,doi:10.1063/1.1637585}.
Including the latter in cold atom quantum simulations would be an interesting topic for further work.

More generally the availability of a larger number of Born-Oppenheimer surfaces in the presence of multiple excitons enlarges the versatility of flexible Rydberg aggregates for the design of tailored BO surfaces, the principles of which are outlined in \cite{wuester:review}. As discussed in that review, an extension to motion in two spatial dimensions typically results in more prominent conical intersections and non-adiabatic dynamics.

\acknowledgments
We gratefully acknowledge fruitful discussions with Alexander Eisfeld, and thank the Science and Engineering Research Board (SERB), Department of Science and Technology (DST), New Delhi, India, for financial support under research Project No.~EMR/2016/005462. Financial support from the Max-Planck society under the MPG-IISER partner group program is also gratefully acknowledged.
\bibliography{multiexciton}

\begin{thebibliography}{37}
\expandafter\ifx\csname natexlab\endcsname\relax\def\natexlab#1{#1}\fi
\expandafter\ifx\csname bibnamefont\endcsname\relax
  \def\bibnamefont#1{#1}\fi
\expandafter\ifx\csname bibfnamefont\endcsname\relax
  \def\bibfnamefont#1{#1}\fi
\expandafter\ifx\csname citenamefont\endcsname\relax
  \def\citenamefont#1{#1}\fi
\expandafter\ifx\csname url\endcsname\relax
  \def\url#1{\texttt{#1}}\fi
\expandafter\ifx\csname urlprefix\endcsname\relax\def\urlprefix{URL }\fi
\providecommand{\bibinfo}[2]{#2}
\providecommand{\eprint}[2][]{\url{#2}}

\bibitem[{\citenamefont{Saffman et~al.}(2010)\citenamefont{Saffman, Walker, and
  M\o{}lmer}}]{Saffman:quantinfryd:review}
\bibinfo{author}{\bibfnamefont{M.}~\bibnamefont{Saffman}},
  \bibinfo{author}{\bibfnamefont{T.~G.} \bibnamefont{Walker}},
  \bibnamefont{and}
  \bibinfo{author}{\bibfnamefont{K.}~\bibnamefont{M\o{}lmer}},
  \bibinfo{journal}{Rev. Mod. Phys.} \textbf{\bibinfo{volume}{82}},
  \bibinfo{pages}{2313} (\bibinfo{year}{2010}).

\bibitem[{\citenamefont{Mohapatra et~al.}(2008)\citenamefont{Mohapatra, Bason,
  Butscher, Weatherill, and Adams}}]{Mohapatra:giantelectroopt}
\bibinfo{author}{\bibfnamefont{A.~K.} \bibnamefont{Mohapatra}},
  \bibinfo{author}{\bibfnamefont{M.~G.} \bibnamefont{Bason}},
  \bibinfo{author}{\bibfnamefont{B.}~\bibnamefont{Butscher}},
  \bibinfo{author}{\bibfnamefont{K.~J.} \bibnamefont{Weatherill}},
  \bibnamefont{and} \bibinfo{author}{\bibfnamefont{C.~S.} \bibnamefont{Adams}},
  \bibinfo{journal}{Nature Physics} \textbf{\bibinfo{volume}{4}},
  \bibinfo{pages}{890} (\bibinfo{year}{2008}).

\bibitem[{\citenamefont{Sevin{\c c}li et~al.}(2011)\citenamefont{Sevin{\c c}li,
  Henkel, Ates, and Pohl}}]{sevincli:nonlocopt}
\bibinfo{author}{\bibfnamefont{S.}~\bibnamefont{Sevin{\c c}li}},
  \bibinfo{author}{\bibfnamefont{N.}~\bibnamefont{Henkel}},
  \bibinfo{author}{\bibfnamefont{C.}~\bibnamefont{Ates}}, \bibnamefont{and}
  \bibinfo{author}{\bibfnamefont{T.}~\bibnamefont{Pohl}},
  \bibinfo{journal}{Phys. Rev. Lett.} \textbf{\bibinfo{volume}{107}},
  \bibinfo{pages}{153001} (\bibinfo{year}{2011}).

\bibitem[{\citenamefont{Peyronel et~al.}(2012)\citenamefont{Peyronel,
  Firstenberg, Liang, Hofferberth, Gorshkov, Pohl, Lukin, and
  Vuleti{\'c}}}]{peyronel:quantnonlinopt}
\bibinfo{author}{\bibfnamefont{T.}~\bibnamefont{Peyronel}},
  \bibinfo{author}{\bibfnamefont{O.}~\bibnamefont{Firstenberg}},
  \bibinfo{author}{\bibfnamefont{Q.-Y.} \bibnamefont{Liang}},
  \bibinfo{author}{\bibfnamefont{S.}~\bibnamefont{Hofferberth}},
  \bibinfo{author}{\bibfnamefont{A.~V.} \bibnamefont{Gorshkov}},
  \bibinfo{author}{\bibfnamefont{T.}~\bibnamefont{Pohl}},
  \bibinfo{author}{\bibfnamefont{M.~D.} \bibnamefont{Lukin}}, \bibnamefont{and}
  \bibinfo{author}{\bibfnamefont{V.}~\bibnamefont{Vuleti{\'c}}},
  \bibinfo{journal}{Nature} \textbf{\bibinfo{volume}{488}}, \bibinfo{pages}{57}
  (\bibinfo{year}{2012}).

\bibitem[{\citenamefont{Dudin and Kuzmich}(2012)}]{dudin:singlephotsource}
\bibinfo{author}{\bibfnamefont{Y.~O.} \bibnamefont{Dudin}} \bibnamefont{and}
  \bibinfo{author}{\bibfnamefont{A.}~\bibnamefont{Kuzmich}},
  \bibinfo{journal}{Science} \textbf{\bibinfo{volume}{336}},
  \bibinfo{pages}{887} (\bibinfo{year}{2012}).

\bibitem[{\citenamefont{Weimer et~al.}(2010)\citenamefont{Weimer, M{\"u}ller,
  Lesanovsky, Zoller, and
  B{\"u}chler}}]{Weimer-Buchler-Rydbergquantumsimulator-2010}
\bibinfo{author}{\bibfnamefont{H.}~\bibnamefont{Weimer}},
  \bibinfo{author}{\bibfnamefont{M.}~\bibnamefont{M{\"u}ller}},
  \bibinfo{author}{\bibfnamefont{I.}~\bibnamefont{Lesanovsky}},
  \bibinfo{author}{\bibfnamefont{P.}~\bibnamefont{Zoller}}, \bibnamefont{and}
  \bibinfo{author}{\bibfnamefont{H.~P.} \bibnamefont{B{\"u}chler}},
  \bibinfo{journal}{Nature Phys.} \textbf{\bibinfo{volume}{6}},
  \bibinfo{pages}{382} (\bibinfo{year}{2010}).

\bibitem[{\citenamefont{Zeiher et~al.}(2016)\citenamefont{Zeiher, van Bijnen,
  Schausz, Hild, Choi, Pohl, Bloch, and Gross}}]{Zeiher:ryd_dress_spinchain}
\bibinfo{author}{\bibfnamefont{J.}~\bibnamefont{Zeiher}},
  \bibinfo{author}{\bibfnamefont{R.}~\bibnamefont{van Bijnen}},
  \bibinfo{author}{\bibfnamefont{P.}~\bibnamefont{Schausz}},
  \bibinfo{author}{\bibfnamefont{S.}~\bibnamefont{Hild}},
  \bibinfo{author}{\bibfnamefont{J.}~\bibnamefont{Choi}},
  \bibinfo{author}{\bibfnamefont{T.}~\bibnamefont{Pohl}},
  \bibinfo{author}{\bibfnamefont{I.}~\bibnamefont{Bloch}}, \bibnamefont{and}
  \bibinfo{author}{\bibfnamefont{C.}~\bibnamefont{Gross}},
  \bibinfo{journal}{Nature Physics} \textbf{\bibinfo{volume}{12}},
  \bibinfo{pages}{1095} (\bibinfo{year}{2016}).

\bibitem[{\citenamefont{Glaetzle et~al.}(2014)\citenamefont{Glaetzle, Dalmonte,
  Nath, Rousochatzakis, Moessner, and Zoller}}]{Glaetzle:spinice:prx}
\bibinfo{author}{\bibfnamefont{A.~W.} \bibnamefont{Glaetzle}},
  \bibinfo{author}{\bibfnamefont{M.}~\bibnamefont{Dalmonte}},
  \bibinfo{author}{\bibfnamefont{R.}~\bibnamefont{Nath}},
  \bibinfo{author}{\bibfnamefont{I.}~\bibnamefont{Rousochatzakis}},
  \bibinfo{author}{\bibfnamefont{R.}~\bibnamefont{Moessner}}, \bibnamefont{and}
  \bibinfo{author}{\bibfnamefont{P.}~\bibnamefont{Zoller}},
  \bibinfo{journal}{Phys. Rev. X} \textbf{\bibinfo{volume}{4}},
  \bibinfo{pages}{041037} (\bibinfo{year}{2014}).

\bibitem[{\citenamefont{Marcuzzi et~al.}(2017)\citenamefont{Marcuzzi,
  Min\'a\ifmmode~\check{r}\else \v{r}\fi{}, Barredo, de~L\'es\'eleuc, Labuhn,
  Lahaye, Browaeys, Levi, and Lesanovsky}}]{Marcuzzi:manybodyloc}
\bibinfo{author}{\bibfnamefont{M.}~\bibnamefont{Marcuzzi}},
  \bibinfo{author}{\bibfnamefont{J.~c.~v.}
  \bibnamefont{Min\'a\ifmmode~\check{r}\else \v{r}\fi{}}},
  \bibinfo{author}{\bibfnamefont{D.}~\bibnamefont{Barredo}},
  \bibinfo{author}{\bibfnamefont{S.}~\bibnamefont{de~L\'es\'eleuc}},
  \bibinfo{author}{\bibfnamefont{H.}~\bibnamefont{Labuhn}},
  \bibinfo{author}{\bibfnamefont{T.}~\bibnamefont{Lahaye}},
  \bibinfo{author}{\bibfnamefont{A.}~\bibnamefont{Browaeys}},
  \bibinfo{author}{\bibfnamefont{E.}~\bibnamefont{Levi}}, \bibnamefont{and}
  \bibinfo{author}{\bibfnamefont{I.}~\bibnamefont{Lesanovsky}},
  \bibinfo{journal}{Phys. Rev. Lett.} \textbf{\bibinfo{volume}{118}},
  \bibinfo{pages}{063606} (\bibinfo{year}{2017}).

\bibitem[{\citenamefont{Wilk et~al.}(2010)\citenamefont{Wilk, Ga\"etan,
  Evellin, Wolters, Miroshnychenko, Grangier, and Browaeys}}]{wilk:entangletwo}
\bibinfo{author}{\bibfnamefont{T.}~\bibnamefont{Wilk}},
  \bibinfo{author}{\bibfnamefont{A.}~\bibnamefont{Ga\"etan}},
  \bibinfo{author}{\bibfnamefont{C.}~\bibnamefont{Evellin}},
  \bibinfo{author}{\bibfnamefont{J.}~\bibnamefont{Wolters}},
  \bibinfo{author}{\bibfnamefont{Y.}~\bibnamefont{Miroshnychenko}},
  \bibinfo{author}{\bibfnamefont{P.}~\bibnamefont{Grangier}}, \bibnamefont{and}
  \bibinfo{author}{\bibfnamefont{A.}~\bibnamefont{Browaeys}},
  \bibinfo{journal}{Phys. Rev. Lett.} \textbf{\bibinfo{volume}{104}},
  \bibinfo{pages}{010502} (\bibinfo{year}{2010}).

\bibitem[{\citenamefont{M\"uller et~al.}(2014)\citenamefont{M\"uller, Murphy,
  Montangero, Calarco, Grangier, and Browaeys}}]{mueller:browaeys:gateoptimise}
\bibinfo{author}{\bibfnamefont{M.~M.} \bibnamefont{M\"uller}},
  \bibinfo{author}{\bibfnamefont{M.}~\bibnamefont{Murphy}},
  \bibinfo{author}{\bibfnamefont{S.}~\bibnamefont{Montangero}},
  \bibinfo{author}{\bibfnamefont{T.}~\bibnamefont{Calarco}},
  \bibinfo{author}{\bibfnamefont{P.}~\bibnamefont{Grangier}}, \bibnamefont{and}
  \bibinfo{author}{\bibfnamefont{A.}~\bibnamefont{Browaeys}},
  \bibinfo{journal}{Phys. Rev. A} \textbf{\bibinfo{volume}{89}},
  \bibinfo{pages}{032334} (\bibinfo{year}{2014}).

\bibitem[{\citenamefont{W{\"u}ster and Rost}(2018)}]{wuester:review}
\bibinfo{author}{\bibfnamefont{S.}~\bibnamefont{W{\"u}ster}} \bibnamefont{and}
  \bibinfo{author}{\bibfnamefont{J.~M.} \bibnamefont{Rost}},
  \bibinfo{journal}{J. Phys. B} \textbf{\bibinfo{volume}{51}},
  \bibinfo{pages}{032001} (\bibinfo{year}{2018}).

\bibitem[{\citenamefont{Gallagher}(1994)}]{book:gallagher}
\bibinfo{author}{\bibfnamefont{T.~F.} \bibnamefont{Gallagher}},
  \emph{\bibinfo{title}{Rydberg Atoms}} (\bibinfo{publisher}{Cambridge
  University Press}, \bibinfo{year}{1994}).

\bibitem[{\citenamefont{Robicheaux et~al.}(2004)\citenamefont{Robicheaux,
  Hernandez, Topcu, and Noordam}}]{noordam:interactions}
\bibinfo{author}{\bibfnamefont{F.}~\bibnamefont{Robicheaux}},
  \bibinfo{author}{\bibfnamefont{J.~V.} \bibnamefont{Hernandez}},
  \bibinfo{author}{\bibfnamefont{T.}~\bibnamefont{Topcu}}, \bibnamefont{and}
  \bibinfo{author}{\bibfnamefont{L.~D.} \bibnamefont{Noordam}},
  \bibinfo{journal}{Phys. Rev. A} \textbf{\bibinfo{volume}{70}},
  \bibinfo{pages}{042703} (\bibinfo{year}{2004}).

\bibitem[{\citenamefont{Ates et~al.}(2008)\citenamefont{Ates, Eisfeld, and
  Rost}}]{cenap:motion}
\bibinfo{author}{\bibfnamefont{C.}~\bibnamefont{Ates}},
  \bibinfo{author}{\bibfnamefont{A.}~\bibnamefont{Eisfeld}}, \bibnamefont{and}
  \bibinfo{author}{\bibfnamefont{J.~M.} \bibnamefont{Rost}},
  \bibinfo{journal}{New J. Phys.} \textbf{\bibinfo{volume}{10}},
  \bibinfo{pages}{045030} (\bibinfo{year}{2008}).

\bibitem[{\citenamefont{W{\"u}ster et~al.}(2010)\citenamefont{W{\"u}ster, Ates,
  Eisfeld, and Rost}}]{wuester:cradle}
\bibinfo{author}{\bibfnamefont{S.}~\bibnamefont{W{\"u}ster}},
  \bibinfo{author}{\bibfnamefont{C.}~\bibnamefont{Ates}},
  \bibinfo{author}{\bibfnamefont{A.}~\bibnamefont{Eisfeld}}, \bibnamefont{and}
  \bibinfo{author}{\bibfnamefont{J.~M.} \bibnamefont{Rost}},
  \bibinfo{journal}{Phys. Rev. Lett.} \textbf{\bibinfo{volume}{105}},
  \bibinfo{pages}{053004} (\bibinfo{year}{2010}).

\bibitem[{\citenamefont{M{\"o}bius et~al.}(2011)\citenamefont{M{\"o}bius,
  W{\"u}ster, Ates, Eisfeld, and Rost}}]{moebius:cradle}
\bibinfo{author}{\bibfnamefont{S.}~\bibnamefont{M{\"o}bius}},
  \bibinfo{author}{\bibfnamefont{S.}~\bibnamefont{W{\"u}ster}},
  \bibinfo{author}{\bibfnamefont{C.}~\bibnamefont{Ates}},
  \bibinfo{author}{\bibfnamefont{A.}~\bibnamefont{Eisfeld}}, \bibnamefont{and}
  \bibinfo{author}{\bibfnamefont{J.~M.} \bibnamefont{Rost}},
  \bibinfo{journal}{J. Phys. B} \textbf{\bibinfo{volume}{44}},
  \bibinfo{pages}{184011} (\bibinfo{year}{2011}).

\bibitem[{\citenamefont{Zoubi et~al.}(2014)\citenamefont{Zoubi, Eisfeld, and
  W{\"u}ster}}]{zoubi:VdWagg}
\bibinfo{author}{\bibfnamefont{H.}~\bibnamefont{Zoubi}},
  \bibinfo{author}{\bibfnamefont{A.}~\bibnamefont{Eisfeld}}, \bibnamefont{and}
  \bibinfo{author}{\bibfnamefont{S.}~\bibnamefont{W{\"u}ster}},
  \bibinfo{journal}{Phys. Rev. A} \textbf{\bibinfo{volume}{89}},
  \bibinfo{pages}{053426} (\bibinfo{year}{2014}).

\bibitem[{\citenamefont{W{\"u}ster et~al.}(2013)\citenamefont{W{\"u}ster,
  M{\"o}bius, Genkin, Eisfeld, and {J.-M. Rost}}}]{wuester:cannon}
\bibinfo{author}{\bibfnamefont{S.}~\bibnamefont{W{\"u}ster}},
  \bibinfo{author}{\bibfnamefont{S.}~\bibnamefont{M{\"o}bius}},
  \bibinfo{author}{\bibfnamefont{M.}~\bibnamefont{Genkin}},
  \bibinfo{author}{\bibfnamefont{A.}~\bibnamefont{Eisfeld}}, \bibnamefont{and}
  \bibinfo{author}{\bibnamefont{{J.-M. Rost}}}, \bibinfo{journal}{Phys. Rev. A}
  \textbf{\bibinfo{volume}{88}}, \bibinfo{pages}{063644}
  (\bibinfo{year}{2013}).

\bibitem[{\citenamefont{van Grondelle et~al.}(1994)\citenamefont{van Grondelle,
  Dekker, Gillbro, and Sundstrom}}]{VANGRONDELLE19941}
\bibinfo{author}{\bibfnamefont{R.}~\bibnamefont{van Grondelle}},
  \bibinfo{author}{\bibfnamefont{J.~P.} \bibnamefont{Dekker}},
  \bibinfo{author}{\bibfnamefont{T.}~\bibnamefont{Gillbro}}, \bibnamefont{and}
  \bibinfo{author}{\bibfnamefont{V.}~\bibnamefont{Sundstrom}},
  \bibinfo{journal}{Biochimica et Biophysica Acta (BBA) - Bioenergetics}
  \textbf{\bibinfo{volume}{1187}}, \bibinfo{pages}{1 } (\bibinfo{year}{1994}).

\bibitem[{\citenamefont{Renger and May}(1997)}]{PhysRevLett.78.3406}
\bibinfo{author}{\bibfnamefont{T.}~\bibnamefont{Renger}} \bibnamefont{and}
  \bibinfo{author}{\bibfnamefont{V.}~\bibnamefont{May}},
  \bibinfo{journal}{Phys. Rev. Lett.} \textbf{\bibinfo{volume}{78}},
  \bibinfo{pages}{3406} (\bibinfo{year}{1997}).

\bibitem[{\citenamefont{Sch{\"o}nleber
  et~al.}(2015)\citenamefont{Sch{\"o}nleber, Eisfeld, Genkin, Whitlock, and
  W{\"u}ster}}]{schoenleber:immag}
\bibinfo{author}{\bibfnamefont{D.~W.} \bibnamefont{Sch{\"o}nleber}},
  \bibinfo{author}{\bibfnamefont{A.}~\bibnamefont{Eisfeld}},
  \bibinfo{author}{\bibfnamefont{M.}~\bibnamefont{Genkin}},
  \bibinfo{author}{\bibfnamefont{S.}~\bibnamefont{Whitlock}}, \bibnamefont{and}
  \bibinfo{author}{\bibfnamefont{S.}~\bibnamefont{W{\"u}ster}},
  \bibinfo{journal}{Phys. Rev. Lett.} \textbf{\bibinfo{volume}{114}},
  \bibinfo{pages}{123005} (\bibinfo{year}{2015}).

\bibitem[{fre()}]{frenkel:footnote}
\bibinfo{note}{Some confusion can arise from the varied use of (Frenkel)
  exciton in the condensed matter, quantum chemistry and molecular aggregate
  community. The former implies an electron-hole pair, with hole localised on a
  crystal site. For a single molecule, and excited state corresponds to a
  Frenkel exciton with hole in the LUMO and electron in the HOMO orbital.
  Finally for molecular aggregates, Frenkel excitons imply the delocalization
  of the latter state over the aggregate due to interactions. Our definition is
  based on the analogy to the latter.}

\bibitem[{\citenamefont{Leonhardt et~al.}(2014)\citenamefont{Leonhardt,
  W{\"u}ster, and Rost}}]{leonhardt:switch}
\bibinfo{author}{\bibfnamefont{K.}~\bibnamefont{Leonhardt}},
  \bibinfo{author}{\bibfnamefont{S.}~\bibnamefont{W{\"u}ster}},
  \bibnamefont{and} \bibinfo{author}{\bibfnamefont{J.~M.} \bibnamefont{Rost}},
  \bibinfo{journal}{Phys. Rev. Lett.} \textbf{\bibinfo{volume}{113}},
  \bibinfo{pages}{223001} (\bibinfo{year}{2014}).

\bibitem[{\citenamefont{Tully}(1990)}]{tully:hopping}
\bibinfo{author}{\bibfnamefont{J.~C.} \bibnamefont{Tully}},
  \bibinfo{journal}{J. Chem. Phys.} \textbf{\bibinfo{volume}{93}},
  \bibinfo{pages}{1061} (\bibinfo{year}{1990}).

\bibitem[{\citenamefont{Tully and Preston}(1971)}]{tully:hopping2}
\bibinfo{author}{\bibfnamefont{J.~C.} \bibnamefont{Tully}} \bibnamefont{and}
  \bibinfo{author}{\bibfnamefont{R.~K.} \bibnamefont{Preston}},
  \bibinfo{journal}{J. Chem. Phys.} \textbf{\bibinfo{volume}{55}},
  \bibinfo{pages}{562} (\bibinfo{year}{1971}).

\bibitem[{\citenamefont{{Hammes-Schiffer} and
  Tully}(1994)}]{tully:hopping:veloadjust}
\bibinfo{author}{\bibfnamefont{S.}~\bibnamefont{{Hammes-Schiffer}}}
  \bibnamefont{and} \bibinfo{author}{\bibfnamefont{J.~C.} \bibnamefont{Tully}},
  \bibinfo{journal}{J. Chem. Phys.} \textbf{\bibinfo{volume}{101}},
  \bibinfo{pages}{4657} (\bibinfo{year}{1994}).

\bibitem[{\citenamefont{Barbatti}(2011)}]{barbatti:review_tully}
\bibinfo{author}{\bibfnamefont{M.}~\bibnamefont{Barbatti}},
  \bibinfo{journal}{Wiley Interdisciplinary Reviews-Computational Molecular
  Science} \textbf{\bibinfo{volume}{1}}, \bibinfo{pages}{620}
  (\bibinfo{year}{2011}).

\bibitem[{\citenamefont{Leonhardt et~al.}(2016)\citenamefont{Leonhardt,
  W{\"u}ster, and Rost}}]{leonhardt:orthogonal}
\bibinfo{author}{\bibfnamefont{K.}~\bibnamefont{Leonhardt}},
  \bibinfo{author}{\bibfnamefont{S.}~\bibnamefont{W{\"u}ster}},
  \bibnamefont{and} \bibinfo{author}{\bibfnamefont{J.~M.} \bibnamefont{Rost}},
  \bibinfo{journal}{Phys. Rev. A} \textbf{\bibinfo{volume}{93}},
  \bibinfo{pages}{022708} (\bibinfo{year}{2016}).

\bibitem[{\citenamefont{Davydov and Kislukha}(1973)}]{davydov:soliton}
\bibinfo{author}{\bibfnamefont{A.~S.} \bibnamefont{Davydov}} \bibnamefont{and}
  \bibinfo{author}{\bibfnamefont{N.~I.} \bibnamefont{Kislukha}},
  \bibinfo{journal}{phys. stat. sol. (b)} \textbf{\bibinfo{volume}{59}},
  \bibinfo{pages}{465} (\bibinfo{year}{1973}).

\bibitem[{\citenamefont{Weidlich and Heudorfer}(1974)}]{weidlich:pulse}
\bibinfo{author}{\bibfnamefont{W.}~\bibnamefont{Weidlich}} \bibnamefont{and}
  \bibinfo{author}{\bibfnamefont{W.}~\bibnamefont{Heudorfer}},
  \bibinfo{journal}{Z. Physik} \textbf{\bibinfo{volume}{268}},
  \bibinfo{pages}{133} (\bibinfo{year}{1974}).

\bibitem[{\citenamefont{Beterov et~al.}(2009)\citenamefont{Beterov, Ryabtsev,
  Tretyakov, and Entin}}]{beterov:BBR}
\bibinfo{author}{\bibfnamefont{I.~I.} \bibnamefont{Beterov}},
  \bibinfo{author}{\bibfnamefont{I.~I.} \bibnamefont{Ryabtsev}},
  \bibinfo{author}{\bibfnamefont{D.~B.} \bibnamefont{Tretyakov}},
  \bibnamefont{and} \bibinfo{author}{\bibfnamefont{V.~M.} \bibnamefont{Entin}},
  \bibinfo{journal}{Phys. Rev. A} \textbf{\bibinfo{volume}{79}},
  \bibinfo{pages}{052504} (\bibinfo{year}{2009}).

\bibitem[{\citenamefont{W{\"u}ster et~al.}(2011)\citenamefont{W{\"u}ster,
  Eisfeld, and Rost}}]{wuester:CI}
\bibinfo{author}{\bibfnamefont{S.}~\bibnamefont{W{\"u}ster}},
  \bibinfo{author}{\bibfnamefont{A.}~\bibnamefont{Eisfeld}}, \bibnamefont{and}
  \bibinfo{author}{\bibfnamefont{J.~M.} \bibnamefont{Rost}},
  \bibinfo{journal}{Phys. Rev. Lett.} \textbf{\bibinfo{volume}{106}},
  \bibinfo{pages}{153002} (\bibinfo{year}{2011}).

\bibitem[{\citenamefont{Leonhardt et~al.}(2017)\citenamefont{Leonhardt,
  W{\"u}ster, and Rost}}]{leonhardt:unconstrained}
\bibinfo{author}{\bibfnamefont{K.}~\bibnamefont{Leonhardt}},
  \bibinfo{author}{\bibfnamefont{S.}~\bibnamefont{W{\"u}ster}},
  \bibnamefont{and} \bibinfo{author}{\bibfnamefont{J.~M.} \bibnamefont{Rost}},
  \bibinfo{journal}{J. Phys. B} \textbf{\bibinfo{volume}{50}},
  \bibinfo{pages}{054001} (\bibinfo{year}{2017}).

\bibitem[{\citenamefont{Saikin et~al.}(2013)\citenamefont{Saikin, Eisfeld,
  Valleau, and {Aspuru-Guzik}}}]{saikin:excitonreview}
\bibinfo{author}{\bibfnamefont{S.~K.} \bibnamefont{Saikin}},
  \bibinfo{author}{\bibfnamefont{A.}~\bibnamefont{Eisfeld}},
  \bibinfo{author}{\bibfnamefont{S.}~\bibnamefont{Valleau}}, \bibnamefont{and}
  \bibinfo{author}{\bibfnamefont{A.}~\bibnamefont{{Aspuru-Guzik}}},
  \bibinfo{journal}{Nanophotonics} \textbf{\bibinfo{volume}{2}},
  \bibinfo{pages}{21} (\bibinfo{year}{2013}).

\bibitem[{\citenamefont{Campillo et~al.}(1976)\citenamefont{Campillo, Shapiro,
  Kollman, Winn, and Hyer}}]{CAMPILLO197693}
\bibinfo{author}{\bibfnamefont{A.}~\bibnamefont{Campillo}},
  \bibinfo{author}{\bibfnamefont{S.}~\bibnamefont{Shapiro}},
  \bibinfo{author}{\bibfnamefont{V.}~\bibnamefont{Kollman}},
  \bibinfo{author}{\bibfnamefont{K.}~\bibnamefont{Winn}}, \bibnamefont{and}
  \bibinfo{author}{\bibfnamefont{R.}~\bibnamefont{Hyer}},
  \bibinfo{journal}{Biophysical Journal} \textbf{\bibinfo{volume}{16}},
  \bibinfo{pages}{93 } (\bibinfo{year}{1976}), ISSN \bibinfo{issn}{0006-3495}.

\bibitem[{\citenamefont{Br{\"u}ggemann and May}(2004)}]{doi:10.1063/1.1637585}
\bibinfo{author}{\bibfnamefont{B.}~\bibnamefont{Br{\"u}ggemann}}
  \bibnamefont{and} \bibinfo{author}{\bibfnamefont{V.}~\bibnamefont{May}},
  \bibinfo{journal}{The Journal of Chemical Physics}
  \textbf{\bibinfo{volume}{120}}, \bibinfo{pages}{2325} (\bibinfo{year}{2004}).

\end{thebibliography}

\end{document}